\documentclass[twocolumn, 9pt]{extarticle}

\usepackage[english]{babel}

\usepackage{amsmath}
\usepackage{graphicx}
\usepackage[colorlinks=true, allcolors=blue]{hyperref}
\usepackage{xcolor}
\usepackage{multirow}
\usepackage{caption}
\usepackage{subcaption}
\usepackage[T1]{fontenc}
\usepackage{cite}
\usepackage{cuted}
\usepackage{capt-of}
\usepackage{dfadobe}  
\usepackage{extsizes}
\usepackage{orcidlink}
\usepackage{abstract}

\title{\huge{Two‑Stage Gaussian Splatting Optimization for Outdoor Scene Reconstruction}}

\author{
  Deborah Pintani\,\orcidlink{0009-0006-0725-729X}$^{1}$ \and
  Ariel Caputo\,\orcidlink{0000-0002-6478-4663}$^{1}$ \and 
  Noah Lewis\,\orcidlink{0009-0000-4446-6994}$^{2}$ \and
  Marc Stamminger\,\orcidlink{0000-0001-8699-3442}$^{2}$ \and
  Fabio Pellacini\,\orcidlink{0000-0003-4861-9809}$^{3}$ \and
  Andrea Giachetti\,\orcidlink{0000-0002-7523-6806}$^{1}$
}

\date{%
  $^1$ University of Verona, Italy \\
  $^2$ Friedrich-Alexander-Universität Erlangen-Nürnberg, Germany \\
  $^3$ University of Modena and Reggio Emilia, Italy
}

\usepackage[a4paper,top=1.5cm,bottom=2cm,left=1.5cm,right=1.5cm]{geometry}
\begin{document}
\maketitle

\begin{strip}
    \centering
    \includegraphics[width=1\linewidth]{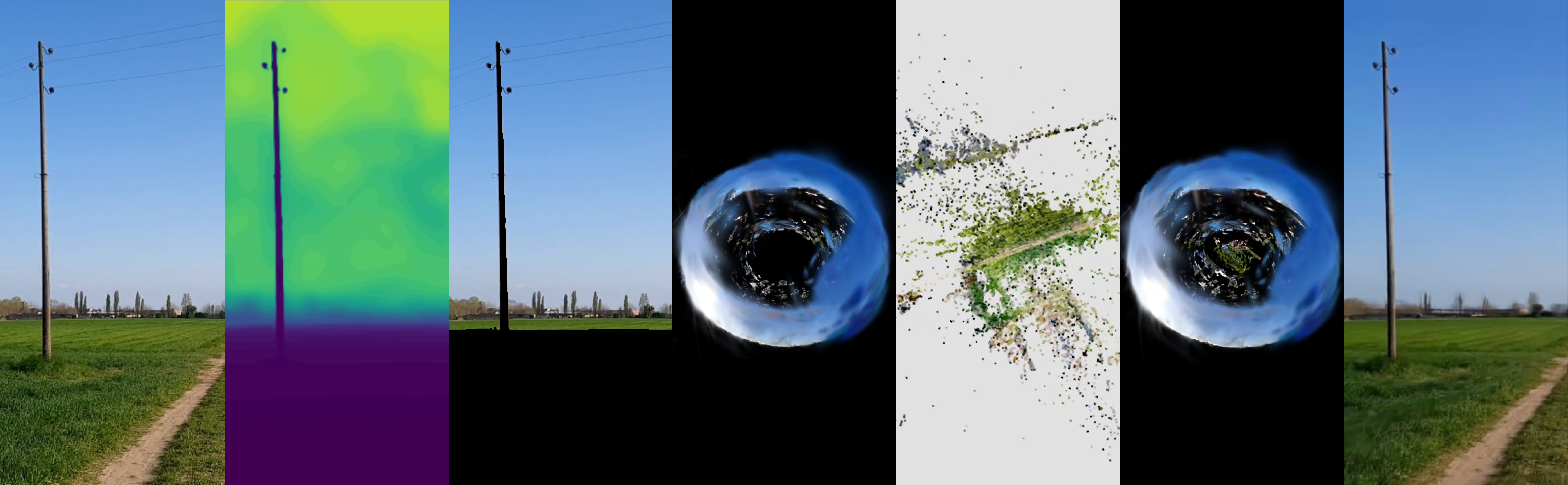}
    \captionof{figure}{All the steps of our two-stage/two-shell Gaussian Splatting pipeline. From left to right: a frame of the input dataset, metric depth map measuring the distance from the center of the scene, masked background image, output of the outer shell's optimization, initialization of the inner shell with photogrammetry, complete reconstruction with outer and inner Gaussian shells, synthetic view.}
    \label{fig:teaser}

    \vspace{1em}
 
    \begin{abstract}
        Outdoor scene reconstruction remains challenging due to the stark contrast between well‑textured, nearby regions and distant backgrounds dominated by low detail, uneven illumination, and sky effects. We introduce a two‑stage Gaussian Splatting framework that explicitly separates and optimizes these regions, yielding higher‑fidelity novel view synthesis.
        In stage one, background primitives are initialized within a spherical shell and optimized using a loss that combines a background‑only photometric term with two geometric regularizers: one constraining Gaussians to remain inside the shell, and another aligning them with local tangential planes. In stage two, foreground Gaussians are initialized from a Structure‑from‑Motion reconstruction, added and refined using the standard rendering loss, while the background set remains fixed but contributes to the final image formation.
        Experiments on diverse outdoor datasets show that our method reduces background artifacts and improves perceptual quality compared to state‑of‑the‑art baselines. Moreover, the explicit background separation enables automatic, object‑free environment map estimation, opening new possibilities for photorealistic outdoor rendering and mixed‑reality applications.
    \end{abstract}
\end{strip}

\section{Introduction}
Gaussian Splatting (GS) \cite{kerbl2023gsplatting} is, nowadays, the most popular approach to encode light-field information from a collection of digital images capturing real-world scenes, enabling real-time rendering of novel views and visual navigation of these scenes. This feature is particularly important to create highly immersive Virtual Reality (VR) applications. The use of Gaussian Splatting for the realization of VR applications is becoming quite popular, with the GS rendering pipeline integrated in game engines like Unreal \footnote{https://dev.epicgames.com/community/learning/tutorials/V2Y2/realityscan-introduction-to-gaussian-splatting-in-unreal-engine} and Unity \footnote{https://github.com/aras-p/UnityGaussianSplatting}, and VR headset manufacturers like Varjo\footnote{https://teleport.varjo.com/} and Meta\footnote{https://www.meta.com/en-gb/experiences/meta-horizon-hyperscape-capture-beta/8798130056953686/} introducing tools for capturing, encoding and rendering GS-based representations of the environments with their devices.

Despite its unquestionable success, the technique presents some critical issues, especially related to the initialization of the Gaussian primitives and their densification and removal strategies.

These issues are particularly relevant in outdoor environments, where it is tough to control the positioning of the Gaussian primitives in regions far from the area where the viewpoints of the input images are located.

Photogrammetry often reconstructs distant objects with only a few, sparse points with large errors. Furthermore, regions corresponding to the sky can suffer from considerable illumination variation or exposure in capture settings. 
These issues can easily lead to the presence of artifacts and floaters. These artifacts are strongly dependent on the initialization and optimization strategy. Many variations of the original approach have been proposed to improve the Gaussians' placement and reduce artifacts, speed up the convergence or reduce the memory usage, but most of them are not designed to reconstruct large outdoor environments. 

In this paper, we present a solution tailored explicitly for reconstructing sets of Gaussians encoding the light field of both close objects and distant background. 

It relies on the segmentation of the scene in specific areas, a \textbf{navigation area}, including the locations where the training images have been captured and where the novel viewpoints that should be generated in the application should lie; a \textbf{nearby area} including the navigation area and a limited neighborhood where it is possible to reconstruct fairly well continuous surfaces of objects with well localized Gaussians, and a \textbf{background area}, for which we represent the related light field with Gaussians that are initialized and constrained on their evolution on a distant sphere and not allowed to grow radially creating floaters. 
The segmentation is based on the distance from the navigation area's center, and it is performed in the training views' image spaces, exploiting dense metric depth maps.

By optimizing the two Gaussian layers sequentially, we obtain a cleaner representation of challenging outdoor environments well suited for VR navigation. 

Our main contributions are:
\begin{itemize} 
    \item A novel framework (Two-Stage/Two-Shell Gaussian Splatting) for outdoor scene reconstruction that explicitly separates foreground and background modeling, creating two concentric shells with Gaussian primitives optimized sequentially with specific procedures.
    \item Two geometric loss functions, $L_{shell}$ and $L_{planarity}$, designed to constrain background Gaussians to a spherical shell and encourage tangential alignment.
    \item A modified pruning strategy that preserves large, stable Gaussians necessary for representing distant scenery
\end{itemize} 

\begin{figure*}[!htb]
    \centering
    \includegraphics[width=1\textwidth]{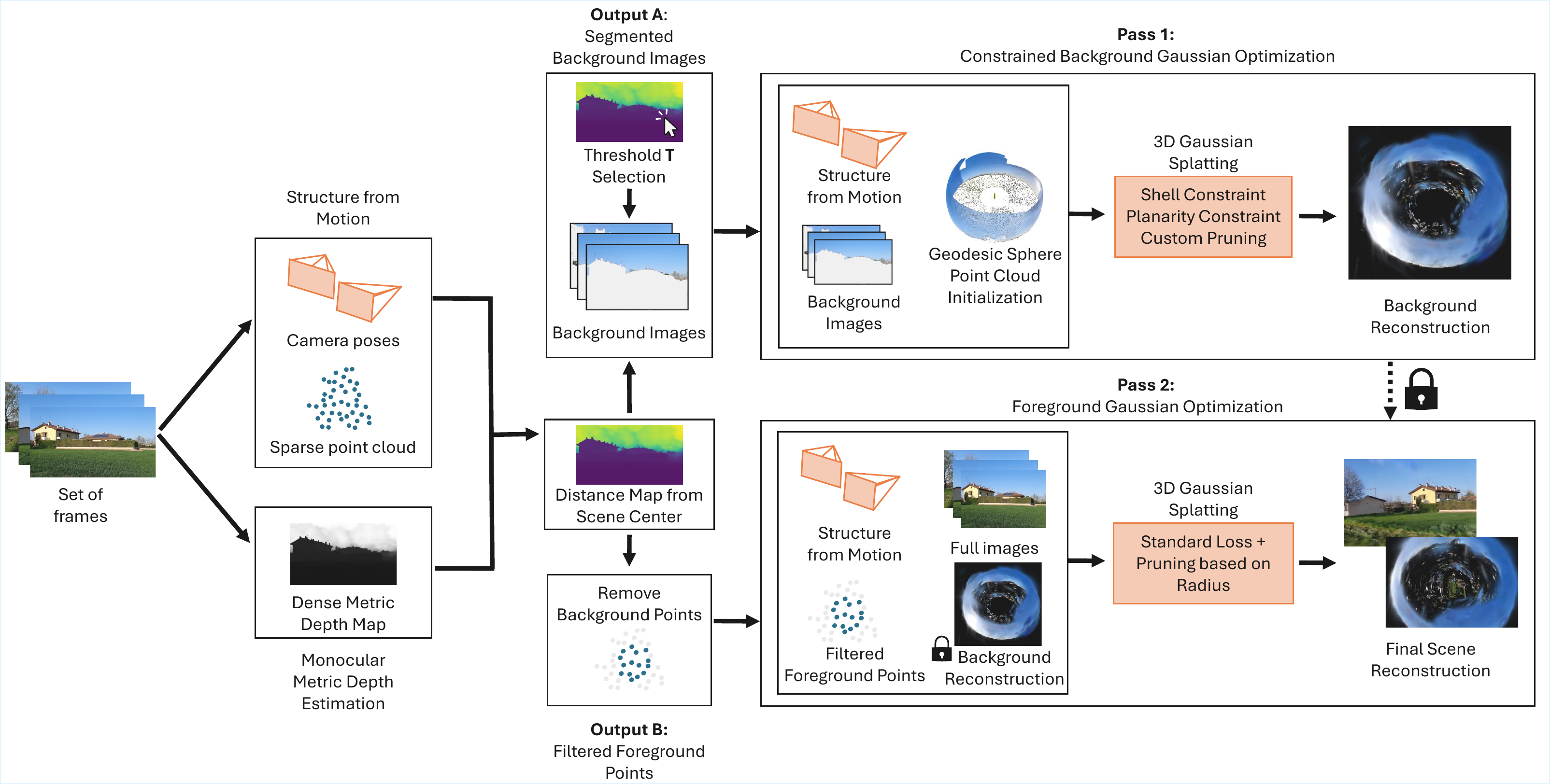}
    \caption{Overview of our two-pass pipeline. From a set of input images, we compute camera poses, a sparse point cloud, and monocular depth maps. Background points are segmented using distance maps and used to initialize a geodesic sphere for the first pass of 3D Gaussian Splatting, producing the background reconstruction with custom losses and pruning. The second pass reconstructs the full scene by combining the background, foreground points, and full images using standard 3D Gaussian Splatting with our pruning strategy.}
    \label{fig:pipeline}
\end{figure*}

We demonstrate with experiments on selected datasets that our method can improve the quality of the novel view synthesis for outdoor environments, especially when the view is far from those sampled in the training data.
We also show that the explicit modeling of the background with the outer shell filled with Gaussians enables the direct estimation of environment maps from the input data.

We will make the code of our Two-Stage/Two-Shell pipeline publicly available soon, including an interactive tool for foreground/background segmentation and the code for environment map extraction.

\begin{figure}[!h]
  \includegraphics[width=1\linewidth]{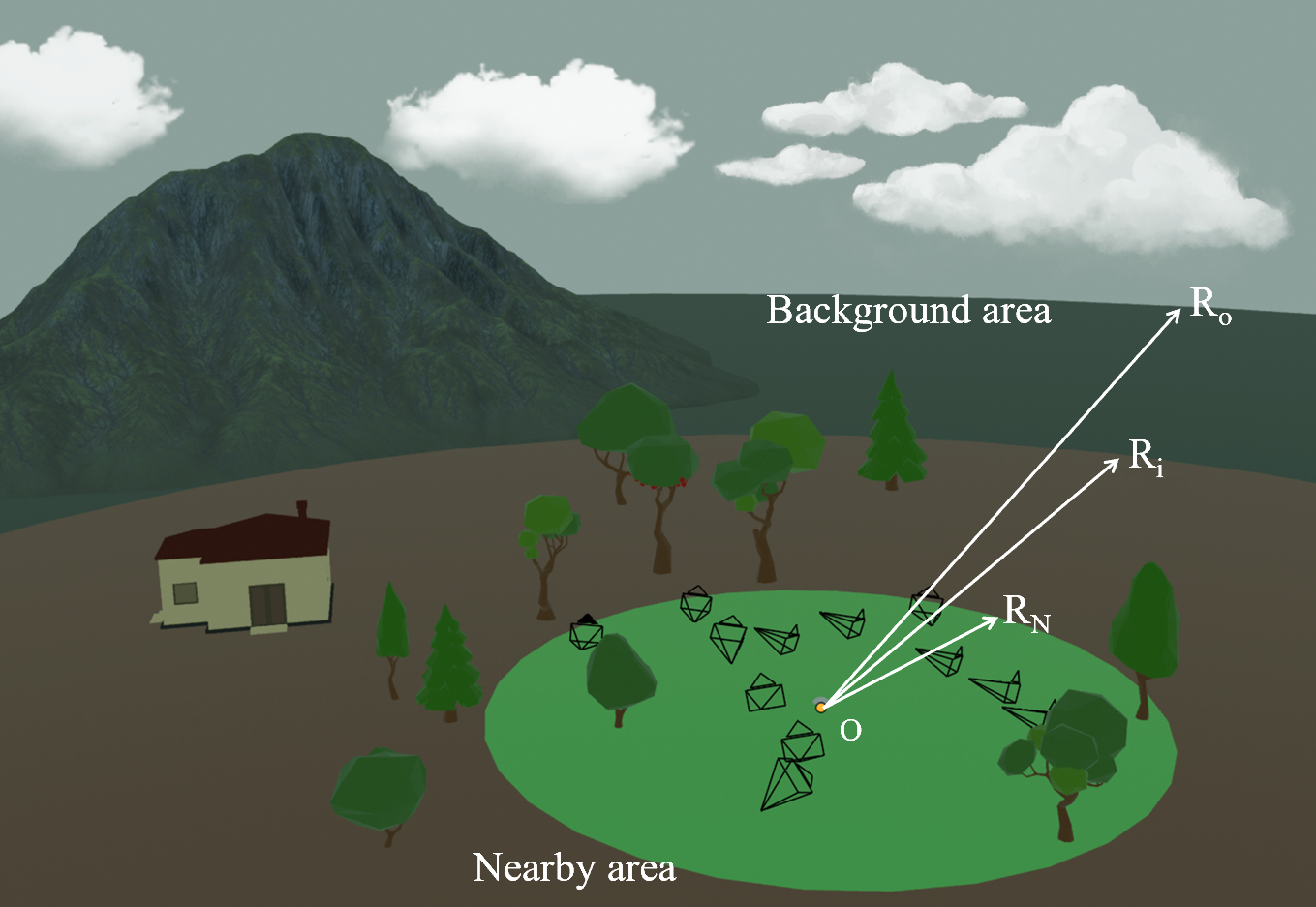}
  \caption{Starting from a collection of photos or a video captured in a limited region (navigation area) of a real environment, we create a Gaussian representation of the light field in the environment with two distinct sets of primitives. The first (background) is constrained to stay within a spherical shell defined by a minimum distance $R_i$ and a maximum distance $R_o$ from the navigation area center. The second (nearby area) represents the objects with a distance from the center lower than $R_i$.}
  \label{fig:sceneRegions}
\end{figure}

\section{Related Works}
Gaussian Splatting (GS) has emerged as a powerful real-time scene representation technique, achieving high-fidelity rendering through optimized anisotropic Gaussian primitives (Kerbl et al. \cite{kerbl2023gsplatting}). Beyond object-centric scenes, recent works have extended GS to handle large-scale and outdoor environments using different approaches to the problem.

Kulhanek et al. identified a relevant issue with outdoor scenes reconstruction, as the Gaussians couldn't be created in the sky with Structure-from-Motion (SfM) initialization \cite{Kulhanek2025}. To overcome this problem, the authors sampled points on a sphere around the scene and added them to the rest of the points used to initialize the Gaussians.
Wang et al. adopted another strategy and implemented a pipeline with different neural modules. They separate the sky from more detailed elements in the scene (e.g., buildings) and generate a cubemap to render the sky \cite{wang2025}. However, while the resulting sky reconstruction looks visually realistic, in some instances (i.e., sky with clouds), it struggles to reproduce images close to the ground truth.

Cheng et al. introduced GaussianPro, using progressive densification strategies to bolster robustness in texture-poor and large outdoor regions \cite{cheng2024gaussianpro}.
Lin et al. proposed VastGaussian, partitioning large scenes into parallel-optimized volumes, allowing real-time splatting across large environments \cite{lin2024vastgaussian}.
Ren et al. developed Octree-GS, employing hierarchical levels of detail (LoD) via octree structures to adaptively manage rendering fidelity in large landscapes \cite{ren2024octreegs}.
In some works, authors tried to tackle the side effects of handling large environments.
Zhang et al. presented GaussianSpa, an optimization-based sparsification framework that reduces point count while preserving visual quality, critical for memory-efficient large-scale GS \cite{zhang2024gaussianspa}.
Pateux et al. designed BOGausS, a better-optimized training regime with confidence-aware updates and rate-distortion densification to enhance GS performance under real-world capture variability \cite{pateux2024bogauss}.
Franke et al. introduced a trilinear point splatting scheme blending GS and neural rendering, showing effective large-scale landscape rendering at real-time rates \cite{franke2024trips}.

Despite these advances, current GS approaches often rely on dense viewpoint coverage, structured capture, or point priors. Modeling distant background regions (e.g., sky, distant mountains, etc.) remains challenging. In Kerbl et al. \cite{Kerbletal2024}, the authors introduce a hierarchy of 3D Gaussians to preserve the visual quality for extensive scenarios. Furthermore, they use a small set of Gaussians to define a spherical skybox just for the sky. With a similar strategy, our method addresses the gaps between nearby and distant elements (including but not limited to the sky) by combining "standard" Gaussians within a defined foreground depth for detailed local fidelity, with a spherical shell of Gaussians (SSG) designed to model the background elements.

\section{Method Description}
    Our pipeline (\autoref{fig:pipeline}) is designed to reconstruct realistic virtual environments from a video (or a sequence of images) captured with a camera moving in a restricted area, called the navigation area.
    It relies on an initial camera pose estimation and scene segmentation in \textbf{nearby} and \textbf{background} areas and a subsequent two-stage optimization to recover the full Gaussian-based representation. This segmentation is shown in \autoref{fig:sceneRegions}.

    \subsection{Pre-processing and Scene Segmentation}

        The first step in our pipeline processes the input images to prepare the data for both optimization stages.

        \subsubsection{Geometric and Depth Estimation}
        
        We begin by running a Structure-from-Motion pipeline (COLMAP \cite{schoenberger2016sfm}) on the input images to obtain camera poses and a 3D sparse point cloud. 
        We place the center of our world system \textbf{O} on the barycenter of the camera positions, although it can be positioned manually in a different place.        
        
        Next, we estimate dense metric depth maps for each image of the input set using a monocular depth estimation method (MMDE). In our experiment, we used Video Depth Anything \cite{video_depth_anything} model, as the 
        datasets used for testing include frames extracted from video streams and the model thus ensures better temporal consistency in the depth estimation across frames. However, several effective MMDE methods are available and can be used instead for generic camera poses \cite{zhang2025survey}. 
        
        Using the camera pose information, we convert the camera-specific depth maps into globally estimated distance maps, where each pixel value represents the metric distance of the corresponding 3D point from the scene origin \textbf{O}.
        
        \subsubsection{Foreground - Background Segmentation.}
        \label{ssec:fg_fb_seg}
        The separation between foreground and background is based on a threshold $R_{i}$ applied to the distance maps. Since this optimal threshold is often scene dependent, we provide an interactive tool to facilitate user-guided segmentation (it will be included in the code distribution). The tool displays a sample of three randomly selected distance maps. They are visualized using a perceptually uniform color map (viridis) to help interpret the distances. By hovering the cursor over different image regions, the user can inspect the metric distance values and determine a suitable threshold $R_{i}$ that effectively divides the background from the foreground. 
        Once confirmed, the threshold is used to produce two assets for the following steps:

        \begin{itemize}
            \item \textbf{Masked Background Images:} The input images are masked to remove all foreground pixels, leaving only the background content for the first optimization stage.
            \item \textbf{Filtered Foreground Points:} The initial sparse point cloud from COLMAP is filtered to keep only the points positioned within the distance threshold $R_{i}$.
        \end{itemize}

        For reproducibility, the selected threshold $R_{i}$ and other scene parameters are written to a file. 

    \subsection{Stage 1: Background Reconstruction}
    \label{sec:stage1}
        \subsubsection{Initialization of the outer Spherical Shell}

        We represent the background as a spherical shell that surrounds the entire scene. We define it as the area between two concentric spheres centered at \textbf{O} (\autoref{fig:sceneRegions}). The smaller one has the radius $R_i$, corresponding to the threshold used to separate the background from the foreground. The radius of the outer one, $R_o$, denotes the maximum limit of the scene, and we set its value to guarantee a bounded parallax when viewed from different cameras placed in the navigation area.
        We empirically found that a value of 10 times the diameter of the navigation area, approximated with the maximum distance between estimated camera positions, is a reasonable tradeoff.
         
        Instead of using sparse points from COLMAP, often unreliable for distant regions, we initialize the background point cloud using a geodesic sphere as a sampling pattern for the polar/azimuthal components. The radial placement of the points is determined using the metric distance maps. If the distance is larger than $R_{o}$, the radius is set precisely to the limit value $R_{o}$, where ideally we place all the points that should be at infinity (e.g., the sky or mountains). If the distance is lower, the radial coordinate of the point is set at a random value in the range [$R_i$ $R_o$]. All the points are initially assigned with view-independent color components equal to the average of the RGB values found in the intersection of the corresponding rays and the input images.
        
\begin{figure}[h]
            \centering
            \includegraphics[width=0.8\linewidth]{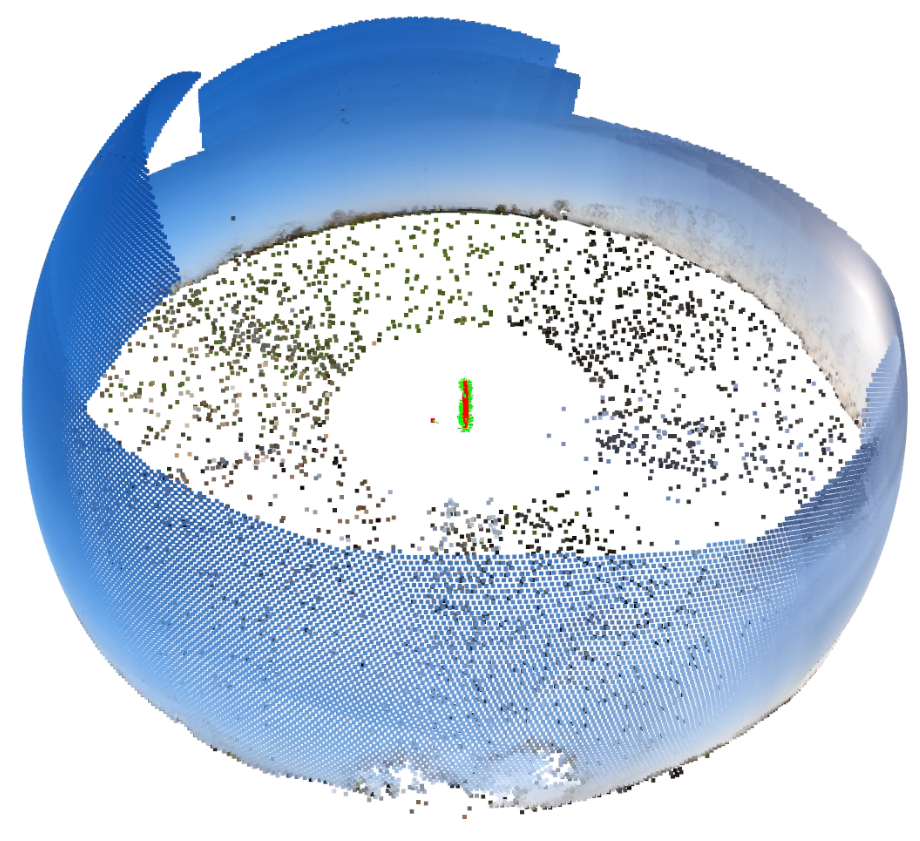}
            \caption{Visualization of the background initialization. Points corresponding to infinity (e.g., the sky) are projected onto the outer sphere in a geodesic pattern, while other distant elements are sampled within the spherical shell.}
            \label{fig:bg_init}
        \end{figure}

        \autoref{fig:bg_init} shows an example initialization. It is worth noting that, as we have the radial depth information, we could use this information for the initial positioning. We tested this choice, but it does not seem to improve the convergence in the outer shell as we show experimentally in Section \ref{sec:abla}.

        \subsubsection{Constrained 3D Gaussian Splatting optimization}
    
        At this step, we perform our customized 3D Gaussian Splatting approach, which reconstructs only the background in this phase. 
        
        The inputs are the segmented background images, the background point cloud generated in the previous stage, the spheres' radii, and the center of the sphere shell.
        During this pass, the Gaussians are constrained to remain inside the spherical shell. To enforce this, we introduce an additional loss function that penalizes any Gaussian whose position falls outside the valid shell, such as closer to the scene center than the inner radius, or farther than the outer radius. 
        The penalty acts as a soft barrier and is proportional to the squared deviation from the shell boundaries, as defined in \autoref{eq:LossShell}:
        
        \begin{equation}
            \label{eq:LossShell}
            L_{shell} = \frac{1}{N} \sum_{i=1}^{N} (max(0, \left\| \textbf{p}_{i} - \textbf{O} \right\|_{2} - R_{o}) + max(0, R_{i} - \left\| \textbf{p}_{i} - \textbf{O} \right\|_{2}))^{2}
        \end{equation}
        
        where $N$ is the total number of Gaussians in the scene, $\textbf{p}_{i}$ is the position of the $i$-th Gaussian, $\textbf{O}$ is the center of the spherical shell, $R_{o}$ is the radius of the outer sphere and $R_{i}$ is the radius of the inner sphere.
        
        We also introduce a planarity loss to discourage the formation of radial spikes, where Gaussians point toward the scene's center. This loss encourages anisotropic Gaussians to lie tangentially on a virtual sphere centered in the spherical shell. Specifically, it penalizes Gaussians whose shortest axis is not aligned with the radial direction from the center to the Gaussian's position. 
        To compute this loss, we first identify the shortest axis in local coordinates for each Gaussian. We then rotate this axis into the global coordinate frame and compute the angle between this axis and the radial vector, calculated from the scene center to the Gaussian's position. The penalty increases as the alignment deviates from the tangency.
        This formula discourages Gaussians from forming radial spikes, and highly anisotropic Gaussians are more penalized, while nearly spherical ones are mostly unaffected. The Equation is shown in \autoref{eq:LossPlanarity}:
        
        \begin{equation}
            \label{eq:LossPlanarity}
            L_{planarity} = \frac{1}{N}\sum_{i=1}^{N}\left[ \left(1- \left| \frac{\textbf{p}_{i} - \textbf{O}}{\left\| \textbf{p}_{i} - \textbf{O} \right\|_{2}} \cdot (Rot_{i} \cdot a_{local, i}) \right| \right) \cdot \frac{s_{max, i}}{s_{min, i} + \epsilon} \right]
        \end{equation}
        
        where $N$ is the total number of Gaussians in the scene, $\textbf{p}_{i}$ is the position of the $i$-th Gaussian, $\textbf{O}$ is the center of the spherical shell, $Rot_{i}$ is the rotation matrix transforming the $i$-th Gaussian from local to global, $a_{local, i}$ is the basis vector in the Gaussian's local coordinate system corresponding to its shortest axis (e.g., $(0, 1, 0)^T$ if the y-axis is the shortest) , $s_{max}$ and $s_{min}$ are the maximum and minimum scaling factors of the $i$-th Gaussian respectively, $\epsilon$ is a small constant to avoid division by zero.
        
        Both the shell loss and the planarity loss are added to the original rendering loss, each weighted by a corresponding $\lambda$ coefficient.

        \subsubsection{Custom Pruning}
        Our custom pipeline also modifies the pruning logic. The original Gaussian Splatting implementation includes a heuristic mechanism aimed at deleting large Gaussians, both in screen space and world space. However, in our case, this strategy can be overly aggressive, particularly for Gaussians representing large and distant background regions. We observed that large Gaussians often modeled these elements, which are incorrectly removed by this strategy. To preserve the integrity and, more importantly, the stability of the background representation, we disable this size-based pruning step. 
        
        On the other hand, we introduce a new pruning strategy that deletes Gaussians that are never seen from any camera viewpoint. The implementation consists of counting the times each Gaussian is rendered, using the indices returned from the rendering step. Any Gaussians with a visibility count equal to zero are removed. This logic is executed in the same pruning stage as in the original implementation.
    
    \subsection{Stage 2: Foreground Reconstruction}

        The second stage of our pipeline focuses on modeling the foreground on top of the already optimized background reconstructed in Section \ref{sec:stage1}.

        \subsubsection{Setup}
    
        We base this reconstruction on the results of the background pass: we load the Gaussians trained in the previous stage, and explicitly include them in the rendering computation to contribute to the final image. However, we keep them fixed during optimization as they are detached from the computation graph. In this way, the background contributes to the rendering but is never updated.

        The foreground is initialized using the filtered point cloud generated during the segmentation step (Section \ref{ssec:fg_fb_seg}).

        An important aspect of this stage is that the foreground reconstruction is confined to the spherical volume defined by the inner radius $R_i$. This radius defines the maximum spatial extent for foreground elements and ensures that the optimization remains focused on the nearby region of interest defined during the initial segmentation.

        \subsubsection{Optimization}

        The foreground Gaussians are optimized using the standard 3D Gaussian Splatting loss function on the original unmasked images. 
        
        To enforce the boundary defined in the setup, we apply a spatial pruning: any foreground Gaussian that moves beyond the inner radius $R_i$ of the background shell is deleted. This constraint ensures a clean separation between the two layers and prevents the foreground from spreading into the background region.

        The final output is a complete scene representation that combines the optimized foreground Gaussians with the separately modeled background.
        
\begin{figure*}[!h]
\captionsetup[subfigure]{labelformat=empty, skip=-8pt}
          \centering
      \begin{subfigure}{0.24\linewidth}
            \includegraphics[width=\linewidth]{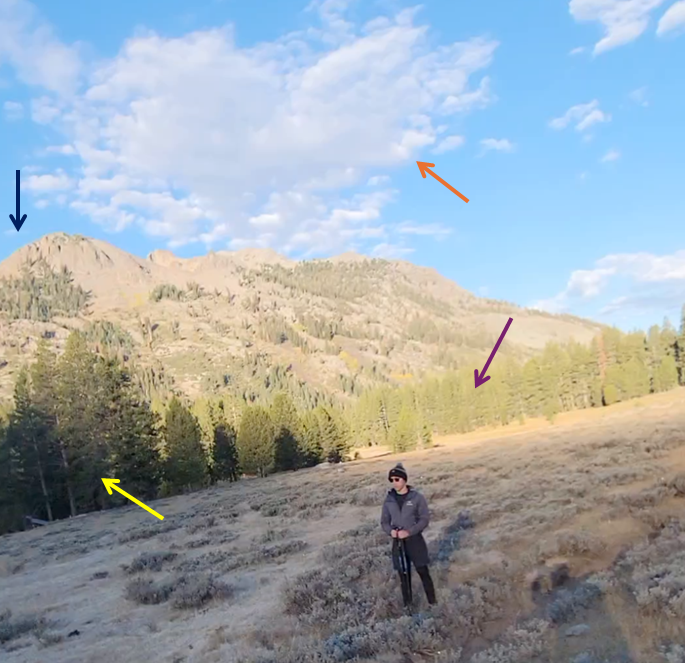}
            \label{fig:personGT}
            \caption{GT}
        \end{subfigure}
      \begin{subfigure}{0.24\linewidth}
            \includegraphics[width=\linewidth]{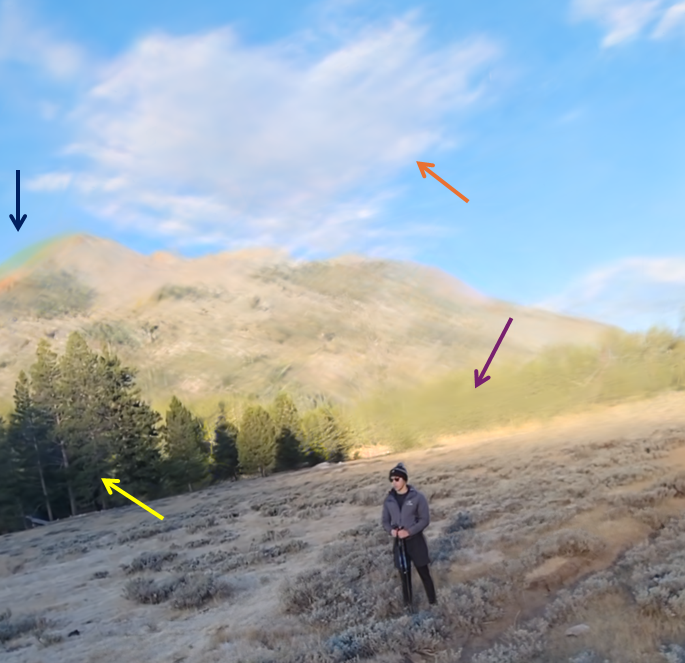}
            \label{fig:personGS}
            \caption{GS (SSIM/LPIPS: 0.852/0.273)}
        \end{subfigure}
      \begin{subfigure}{0.24\linewidth}
            \includegraphics[width=\linewidth]{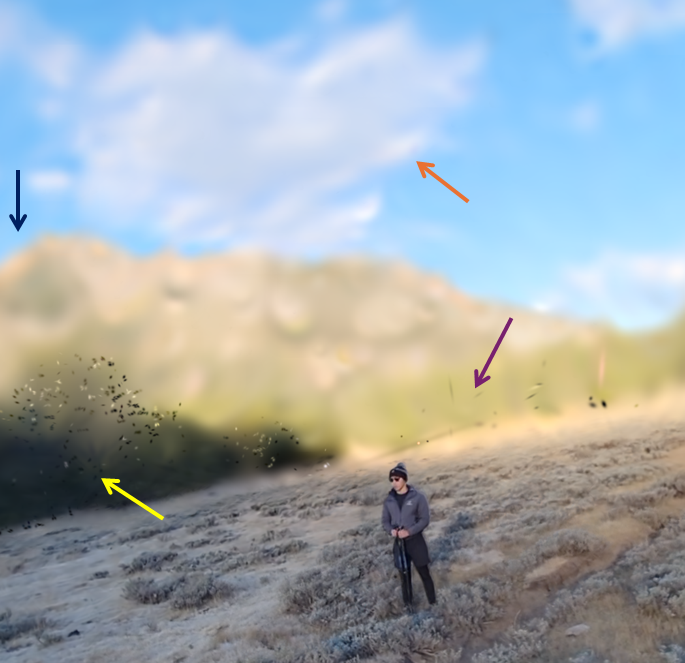}
            \label{fig:personHGS}
            \caption{HGS  (SSIM/LPIPS: 0.842/0.288)}
        \end{subfigure}
      \begin{subfigure}{0.24\linewidth}
            \includegraphics[width=\linewidth]{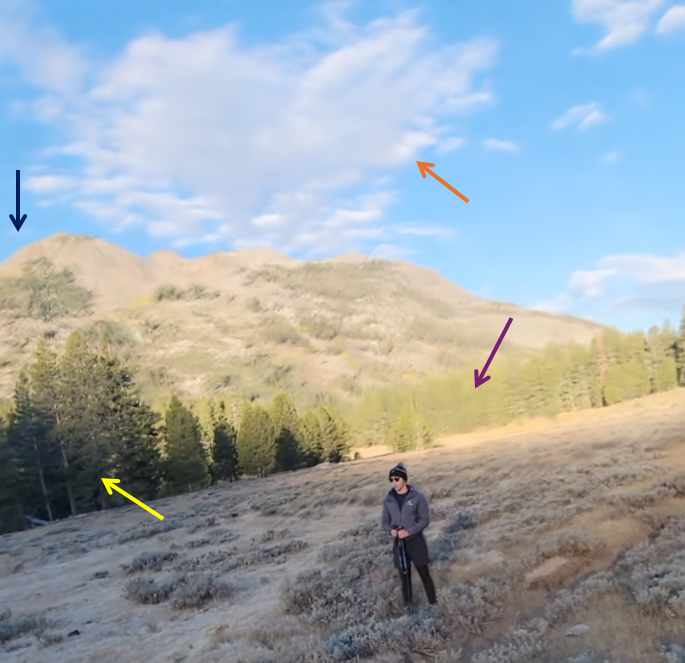}
            \label{fig:personOURS}
            \caption{Ours  (SSIM/LPIPS: 0.865/0.244)}
        \end{subfigure}

      \begin{subfigure}{0.24\linewidth}
            \includegraphics[width=\linewidth]{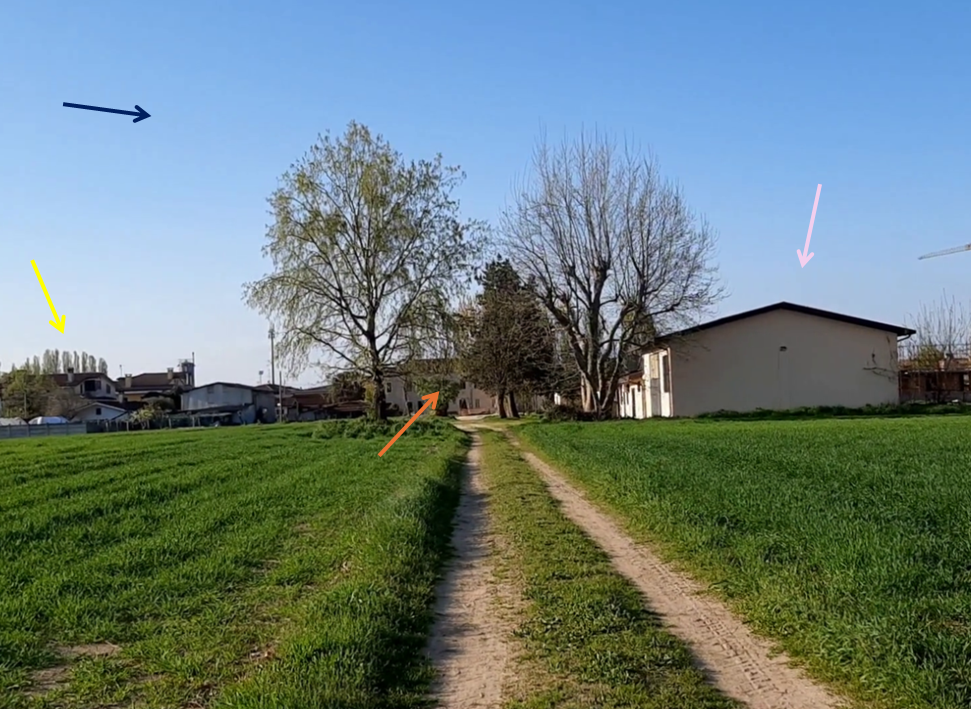}
            \label{fig:fieldsGT}
            \caption{GT}
        \end{subfigure}
      \begin{subfigure}{0.24\linewidth}
            \includegraphics[width=\linewidth]{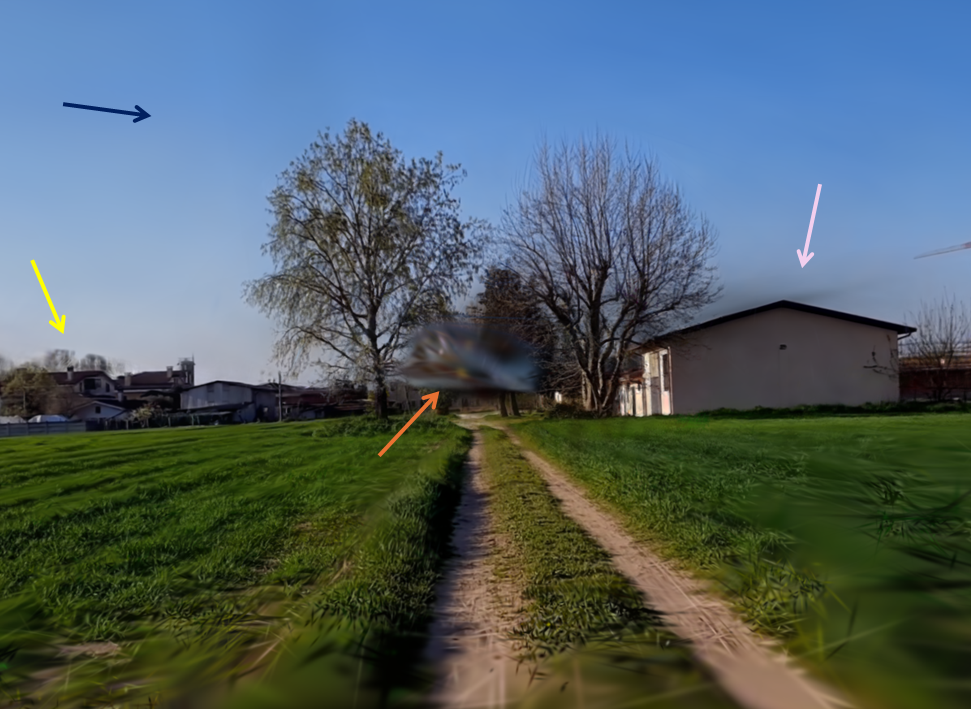}
            \label{fig:fieldsGS}
            \caption{GS (SSIM/LPIPS: 0.667/0.273)}
        \end{subfigure}
      \begin{subfigure}{0.24\linewidth}
            \includegraphics[width=\linewidth]{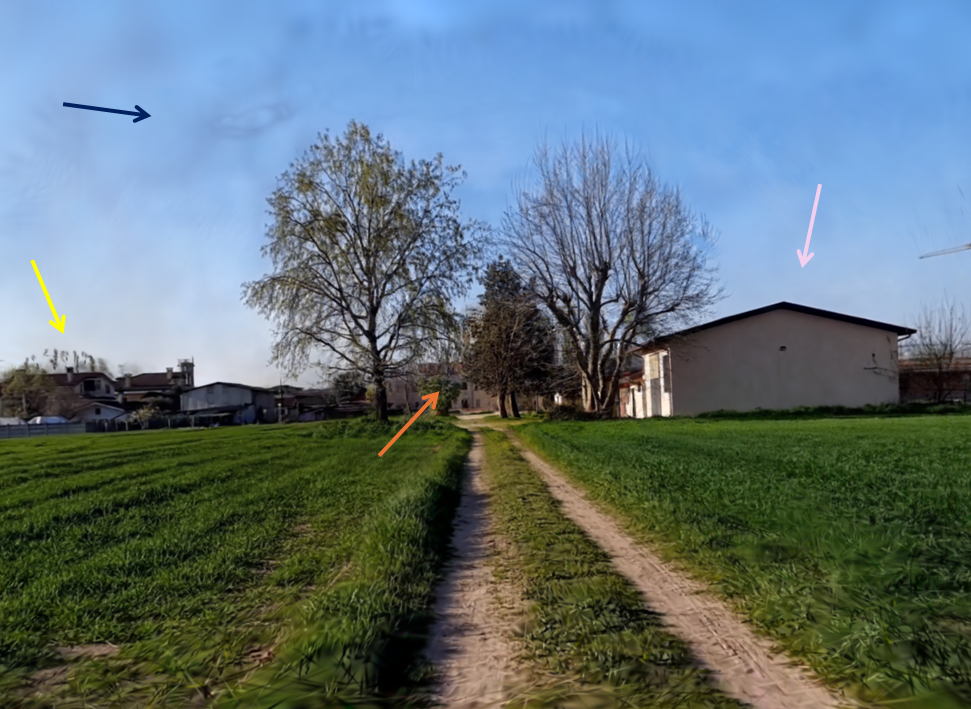}
            \label{fig:fieldsHGS}
            \caption{HGS (SSIM/LPIPS: 0.713/0.249)}
        \end{subfigure}
      \begin{subfigure}{0.24\linewidth}
            \includegraphics[width=\linewidth]{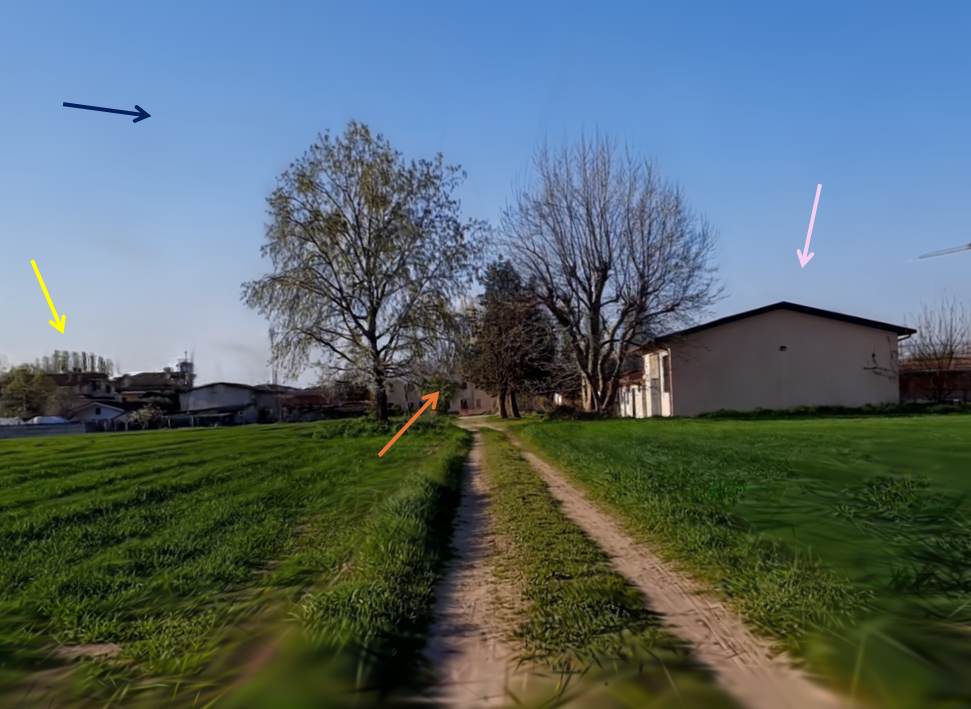}
            \label{fig:fieldsOURS}
            \caption{Ours (SSIM/LPIPS: 0.693/0.255)}
        \end{subfigure}

              \begin{subfigure}{0.24\linewidth}
            \includegraphics[width=\linewidth]{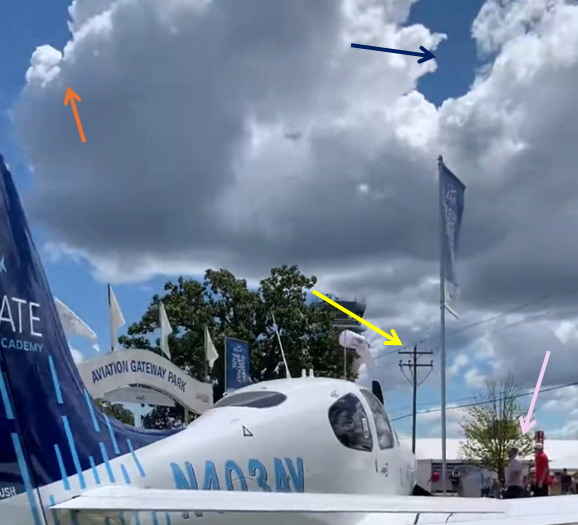}
            \label{fig:planeGT}
            \caption{GT}
        \end{subfigure}
      \begin{subfigure}{0.24\linewidth}
            \includegraphics[width=\linewidth]{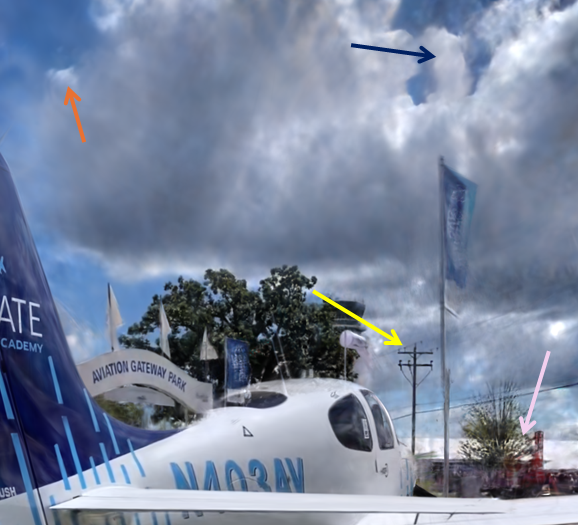}
            \label{fig:planeGS}
            \caption{GS (SSIM/LPIPS: 0.697/0.346)}
        \end{subfigure}
      \begin{subfigure}{0.24\linewidth}
            \includegraphics[width=\linewidth]{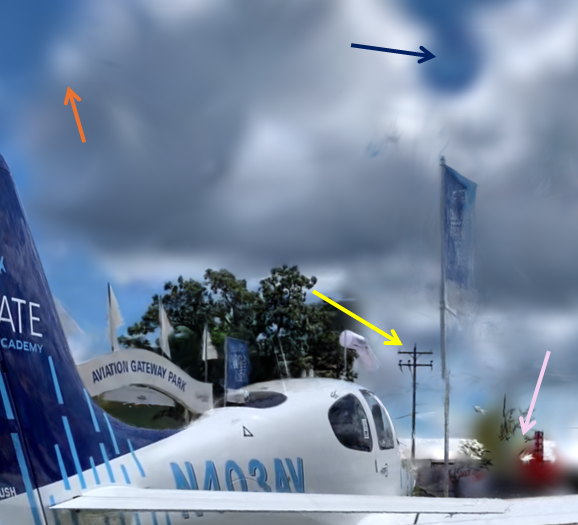}
            \label{fig:planeHGS}
            \caption{HGS (SSIM/LPIPS: 0.697/0.316)}
        \end{subfigure}
      \begin{subfigure}{0.24\linewidth}
            \includegraphics[width=\linewidth]{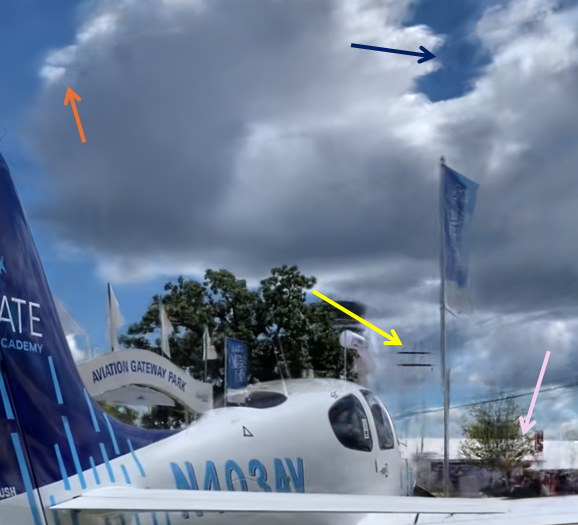}
            \label{fig:planeOURS}
            \caption{Ours (SSIM/LPIPS: 0.706/0.327)}
        \end{subfigure}

              \begin{subfigure}{0.24\linewidth}
            \includegraphics[width=\linewidth]{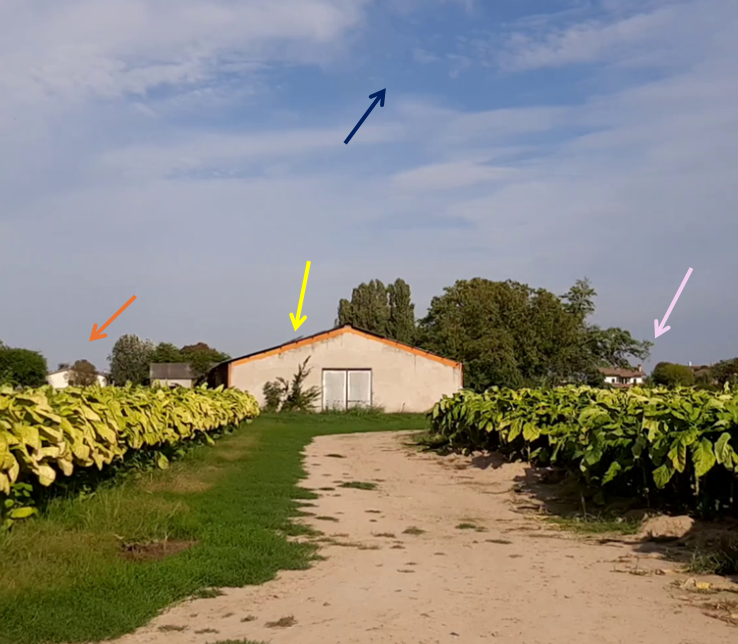}
            \label{fig:tobaccoGT}
            \caption{GT}
        \end{subfigure}
      \begin{subfigure}{0.24\linewidth}
            \includegraphics[width=\linewidth]{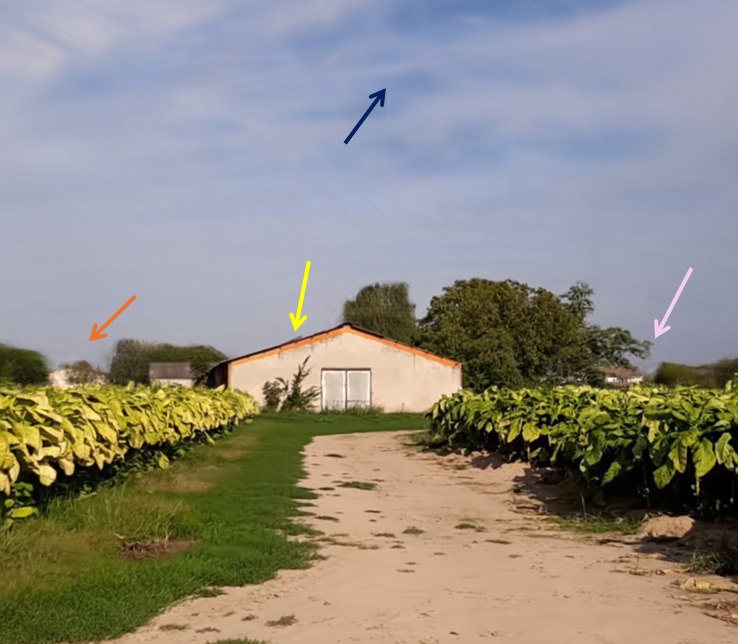}
            \label{fig:tobaccoGS}
            \caption{GS (SSIM/LPIPS: 0.873/0.165)}
        \end{subfigure}
      \begin{subfigure}{0.24\linewidth}
            \includegraphics[width=\linewidth]{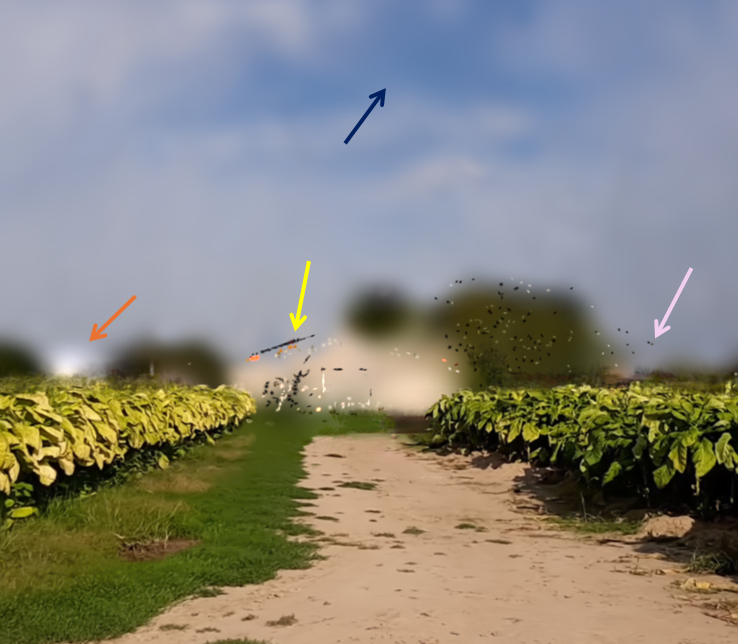}
            \label{fig:tobaccoHGS}
            \caption{HGS (SSIM/LPIPS: 0.853/0.203)}
        \end{subfigure}
      \begin{subfigure}{0.24\linewidth}
            \includegraphics[width=\linewidth]{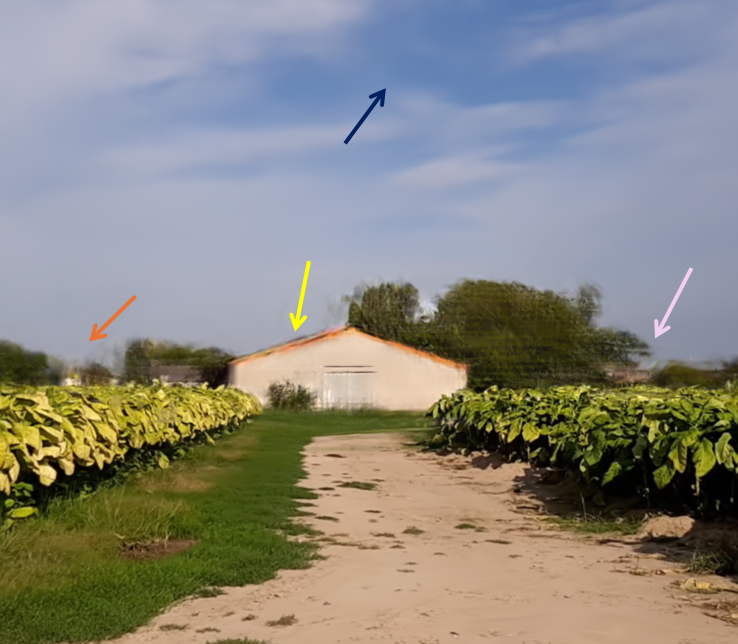}
            \label{fig:tobaccoOURS}
            \caption{Ours (SSIM/LPIPS: 0.853/0.177)}
        \end{subfigure}

              \begin{subfigure}{0.24\linewidth}
            \includegraphics[width=\linewidth]{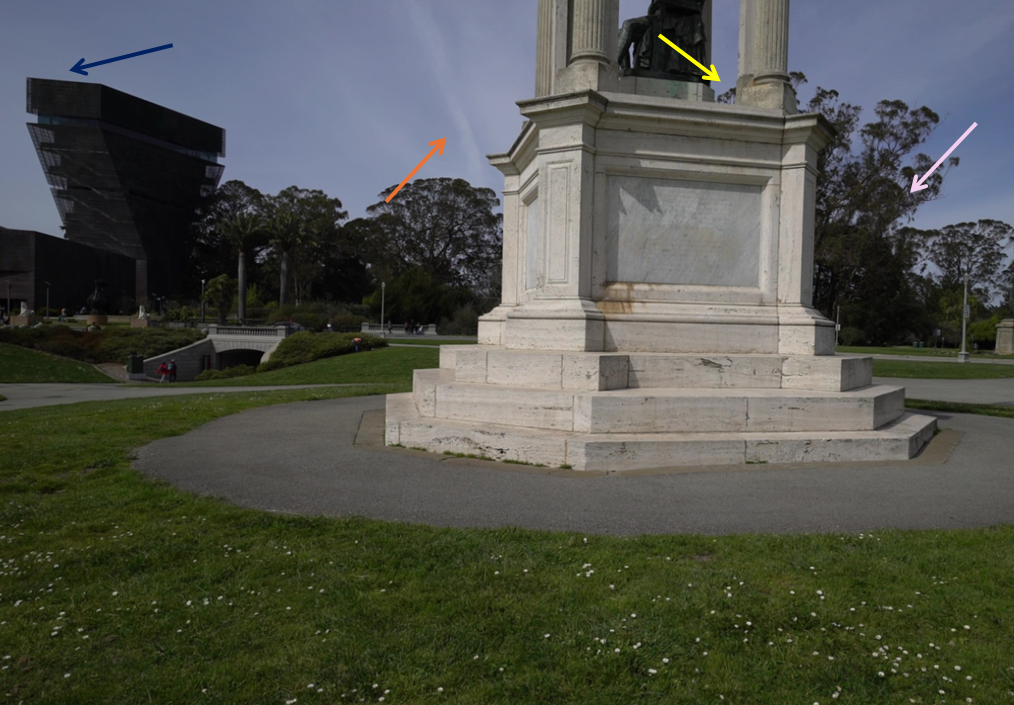}
            \label{fig:francisGT}
            \caption{GT}
        \end{subfigure}
      \begin{subfigure}{0.24\linewidth}
            \includegraphics[width=\linewidth]{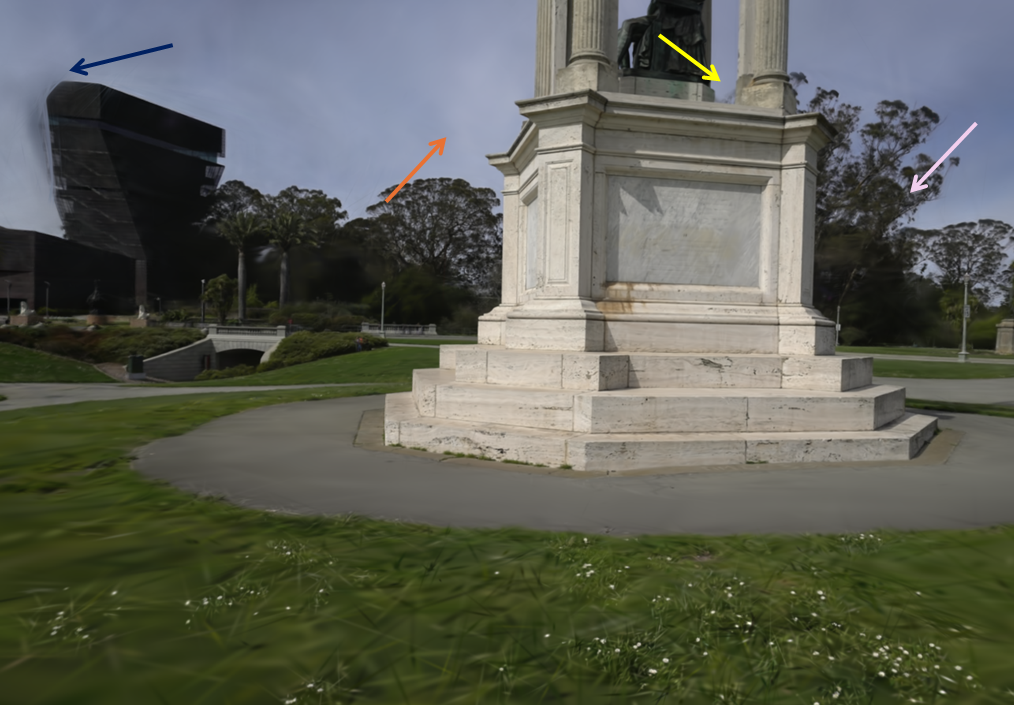}
            \label{fig:francisGS}
            \caption{GS (SSIM/LPIPS: 0.849/0.258)}
        \end{subfigure}
      \begin{subfigure}{0.24\linewidth}
            \includegraphics[width=\linewidth]{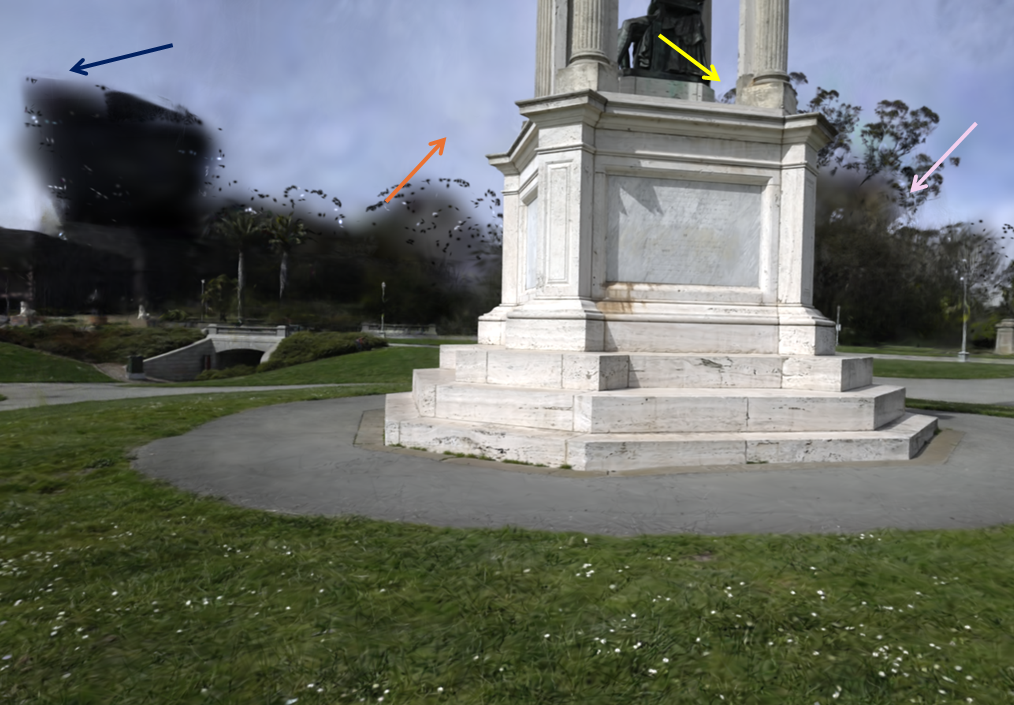}
            \label{fig:francisHGS}
            \caption{HGS (SSIM/LPIPS: 0.804/0.260)}
        \end{subfigure}
      \begin{subfigure}{0.24\linewidth}
            \includegraphics[width=\linewidth]{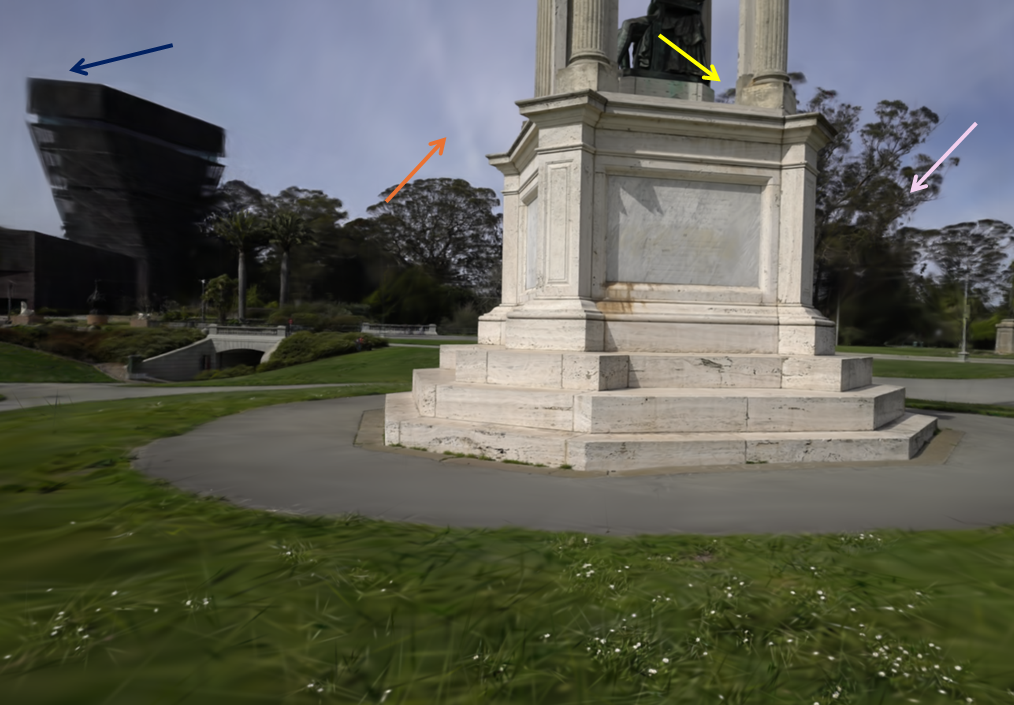}
            \label{fig:francisOURS}
            \caption{Ours (SSIM/LPIPS: 0.852/0.264)}
        \end{subfigure}

          \caption{Comparison of novel view syntheses for the five benchmark datasets. Arrows highlight regions with relevant artifacts (see text). Our method consistently provides artifact-free results for sky and very distant objects.}
          \label{fig:nvs}
\end{figure*}

\section{Experimental Results}

We evaluated our framework on five outdoor datasets, selected to test its capabilities. The choice focused on scenes with the following properties:

\begin{itemize}
    \item A relatively constrained camera navigation area
    \item A depth separation between a nearby foreground region and a distant, open background containing a significant portion of elements like sky, faraway buildings, or trees.
\end{itemize}

These specifics are typical of object-centric captures, where the camera moves around an object, but are not limited to them: we also included more general landscape scenes where a separation exists between the immediate surroundings and the background. 
The datasets chosen are: 

\begin{itemize}
    \item \textit{Person} and \textit{Plane} from NerfStudio \cite{Tancik_2023}
    \item \textit{Francis} from Tanks and Temples \cite{Knapitsch2017}
    \item \textit{Fields} and \textit{Tobacco}, specifically captured by us. 
\end{itemize}

Both \textit{Fields} and \textit{Tobacco} are videos recorded with a smartphone camera, from which we extracted frames at 2 FPS. In particular, \textit{Fields} contains a total of 602 frames, while \textit{Tobacco} has 141 extracted frames.

For our specific novel view synthesis benchmarking task, each image sequence was randomly divided into a training set (80\% of the frames) and a test set (20\%).

For each dataset, we optimized the Gaussian environments using our pipeline, the official GS code \cite{kerbl2023gsplatting}, and the official Hierarchical GS code (HGS) \cite{Kerbletal2024}. Regarding HGS, we decided to use the maximum quality parameters, corresponding to $\tau = 0$.

In all experiments, exposure compensation was enabled during training. Following standard practice, test images were not provided as input, meaning the compensation was computed only on the training images.
Furthermore, we ensured that all methods were initialized with the same point cloud for the training of the nearby area.

We ran all experiments on a machine mounting an NVIDIA RTX 5090 (32 GB), an AMD Ryzen 9 7900X CPU, and 96 GB of RAM. On this machine, our method's training time was approximately 40-50 minutes for both the stages.

\autoref{tab:metrics} shows the average similarity metrics comparing the ground truth test images with the corresponding ones estimated with the three light field encodings.
Our method provided average values that improved the current version of the official GS code and the hierarchical GS version without adding complexity regarding the number of Gaussian primitives employed. 

However, looking at the example novel views synthesized with our method  (\autoref{fig:nvs}), and visually comparing them with the results of traditional and hierarchical GS optimization, it is possible to see that we achieve strong improvements in the quality of the sky and of the appearance of distant objects. 

\begin{table}[]
\begin{tabular}{cl|lll}
\multicolumn{1}{l}{Scene}    &       & \multicolumn{3}{c}{Method}        \\
\multicolumn{1}{l}{}         &       & Ours    & 3DGS    & H3DGS ($\tau = 0$) \\ \hline
\multirow{4}{*}{Person}      & SSIM $\uparrow$  & 0,831    & 0,813    & 0,836          \\
                             & PSNR  $\uparrow$& 29,242   & 28,045   & 26,835         \\
                             & LPIPS $\downarrow$& 0,230    & 0,252    & 0,249          \\
                             & N°    & 1709832 & 1444442 & 1903744       \\ \hline
\multirow{4}{*}{Fields}      & SSIM  $\uparrow$& 0,750    & 0,740    & 0,738          \\
                             & PSNR  $\uparrow$& 23,291   & 22,459   & 22,394         \\
                             & LPIPS $\downarrow$ & 0,231    & 0,224    & 0,297          \\
                             & N°    & 2941561 & 2758201 & 2291674       \\ \hline
\multirow{4}{*}{Plane}       & SSIM $\uparrow$ & 0,723    & 0,713    & 0,707          \\
                             & PSNR $\uparrow$ & 21,368   & 20,667   & 20,637         \\
                             & LPIPS $\downarrow$& 0,311    & 0,322    & 0,326          \\
                             & N°    & 1758114 & 1756838 & 3135991       \\ \hline
\multirow{4}{*}{Tobacco}     & SSIM $\uparrow$ & 0,776    & 0,769    & 0,738          \\
                             & PSNR $\uparrow$ & 24,007   & 23,058   & 23,317         \\
                             & LPIPS $\downarrow$ & 0,215    & 0,218    & 0,259          \\
                             & N°    & 2758527 & 2776583 & 1309581       \\ \hline
\multirow{4}{*}{Francis}     & SSIM $\uparrow$ & 0,883    & 0,889    & 0,834          \\
                             & PSNR $\uparrow$ & 22,765   & 22,770   & 17,206         \\
                             & LPIPS $\downarrow$& 0,227    & 0,219    & 0,272          \\
                             & N°    & 786195  & 649235  & 1218190   
                             \\ \hline
\multirow{4}{*}{\textbf{Average}}     & SSIM $\uparrow$ & 0,784    & 0,788   & 0,767         \\
                             & PSNR $\uparrow$ & 24,156   & 23,499   & 22,428           \\
                             & LPIPS $\downarrow$& 0,244  & 0,252  & 0,259            \\
                             & N°    & 1990846  & 1877060  & 1971836

                              \\ \hline
\end{tabular}
\caption{Quantitative evaluation (average similarity scores of comparisons between test views and synthesized views) of our method compared to 3DGS and the hierarchical 3DGS official implementations.}
\label{tab:metrics}
\end{table}

\begin{figure}[h]
  \centering
  \begin{subfigure}{0.49\linewidth}
    \includegraphics[width=\linewidth]{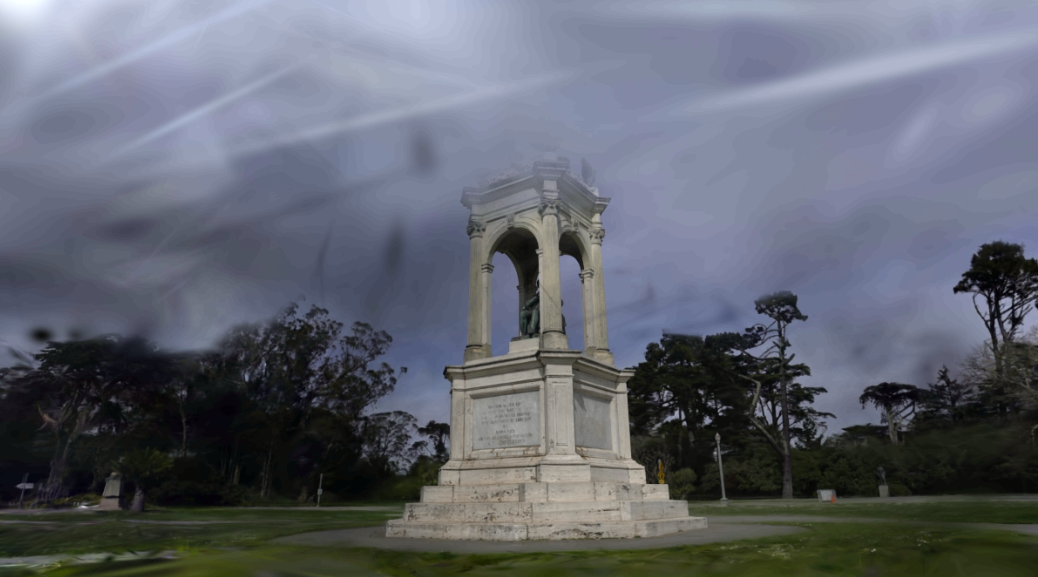}
    \caption{GS, novel view}
    \label{fig:francis_gs}
  \end{subfigure}\hfill
  \begin{subfigure}{0.49\linewidth}
    \includegraphics[width=\linewidth]{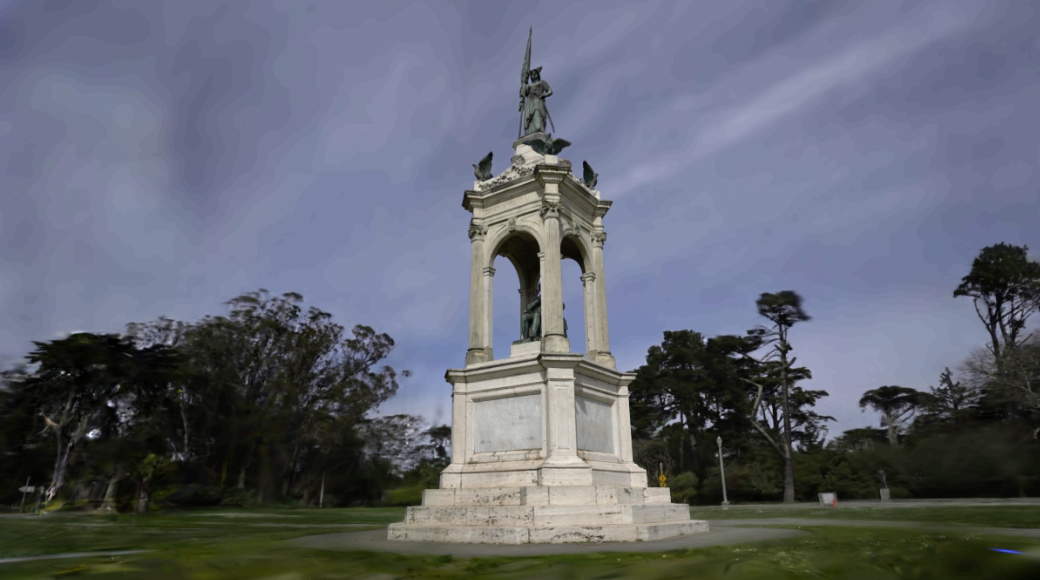}
    \caption{Ours, novel view}
    \label{fig:francis_ours}
  \end{subfigure}

  \begin{subfigure}{0.49\linewidth}
    \includegraphics[width=\linewidth]{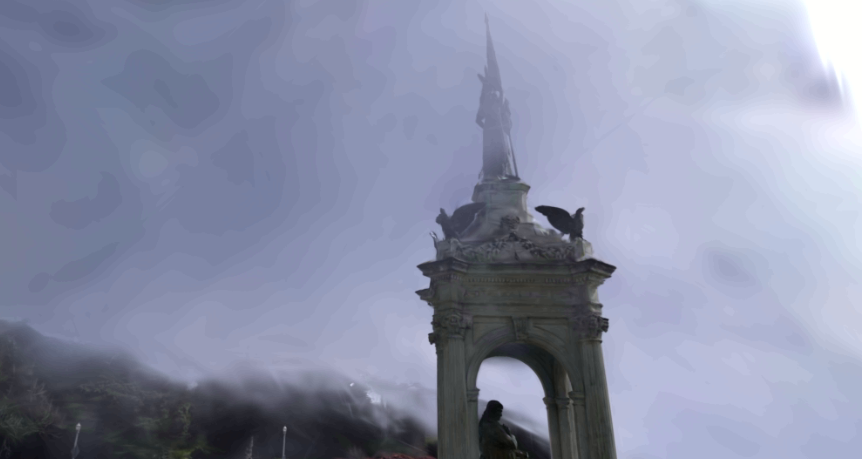}
    \caption{GS, novel view}
    \label{fig:francis_gs2}
  \end{subfigure}\hfill
  \begin{subfigure}{0.49\linewidth}
    \includegraphics[width=\linewidth]{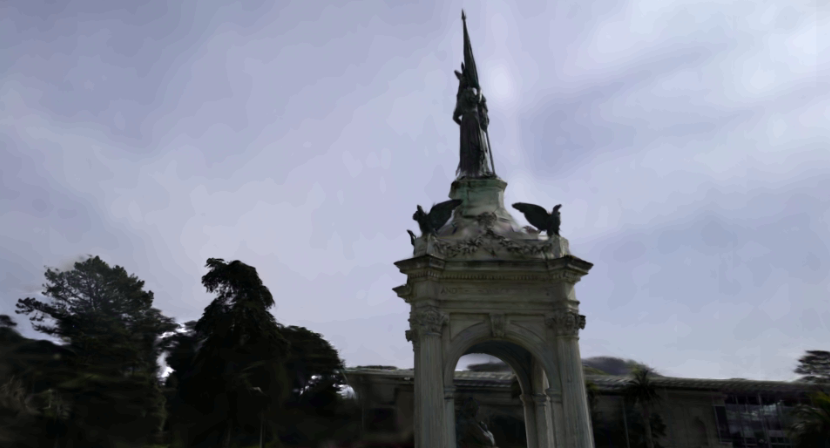}
    \caption{Ours, novel view}
    \label{fig:francis_ours2}
  \end{subfigure}

  \caption{Comparison of two novel views generated from the Gaussian distributions trained on the \textit{Francis} datasets and rendered from cameras with parameters far from those included in the training set. Standard GS renderings are affected by the presence of elongated primitives, which obscure the scene elements. Our solution provides artifact-free rendering.}
  \label{fig:francis}
\end{figure}

In the \textit{Person} example (first row), our technique results in sharp clouds similar to the GT ones (orange arrow). GS also shows chromatic artifacts on the top of the mountain (dark blue arrow). Our method also provides a sharper reconstruction of the trees (yellow and purple arrows). Looking at the trees on the left, we see that HGS creates a layered image with sharp elements popping up from a smooth background (see yellow arrows).

In the example from the \textit{Fields} dataset (second row), our method provides a clean sky as the original GS. At the same time, the hierarchical GS shows evident artifacts with posterized regions (blue arrows). GS presents evident artifacts due to floaters, a white region in front of the central tree (orange arrow), and an apparent leakage of the color of the roof (pink arrow). Our method is also more accurate in reconstructing the trees on the left (yellow arrows). 

In the example from the \textit{Plane} dataset (third row), our method also provides the only sky similar to the original one, as HGS creates oversmoothed clouds and GS adds non-existing structures (see orange and blue arrows). In this case, however, our method produces more artifacts in some closer details (yellow and pink arrows) 

In the examples from the \textit{Tobacco} dataset (fourth row), our method is again the one creating a sky similar to the GT (blue arrow), with still GS creating non-existing patterns and HGS oversmoothing. However, in this case, our method provides more artifacts than the original GS method near the horizon (orange, yellow, and pink arrows).

Finally, in the examples from the \textit{Francis} dataset (bottom row), the advantages of our technique are evident: it is the only creating a correct visualization of the building on the left (blue arrow), the proper pattern of the trail in the sky (orange line), disappearing in the others, a correct reprodiction of the tree behind the statue (yellow arrow). Here, the hierarchical GS also creates a pop-up effect with sharp branches of the tree popping up from a blurred background (pink arrow).

It is worth noting that the evident artifacts discussed, especially those in the sky, do not penalize the evaluation metrics used to compare the different approaches too much. Overblurred clouds and trees of HGS in \textit{Tobacco} do not make SSIM too bad. The dark spots in the sky in the \textit{Fields} image seem not to affect LPIPS as well. Even the bad artifacts in the building in the \textit{Francis} images estimated with GS and HGS do not seem to affect the LPIPS metrics, which appear better than ours on the selected image, even if our result appears clearly the most realistic.

This suggests that the evaluation of the quality of the virtual environments reconstructed with light fields representation should not rely only on the quantitative comparison with the standard metrics and that different methods should be designed for specific applications (e.g., user studies).

Another evident limitation in evaluating the light fields' quality based on the quantitative tests on test views extracted from the data acquired in the typical video capture of these datasets, is that, with this choice, the test camera poses are not too different to some of the training ones. 
This makes typical issues of 3DGS, like the presence of floaters in areas that should be empty, difficult to see. 
However, for the application of our interest (e.g., navigation in VR), it is not sufficient that the synthesized view are good for camera poses similar to the training ones, but need to be suitable for \textit{any camera pose in the "navigation" area} where the user may want to go exploring the scene (\autoref{fig:sceneRegions}).
Thanks to the specific design of the optimization, our method can guarantee floater-free scenes in all the nearby areas, with strong improvements for the user immersion. 
\autoref{fig:francis} shows clearly this effect: while the selection of a camera with position and orientations not similar to the training one makes the standard GS renderings affected by floaters (a,c), our results are clean and realistic (b,d).

   \begin{figure*}[ht]
          \centering
          \begin{subfigure}{0.32\linewidth}
            \includegraphics[width=\linewidth]{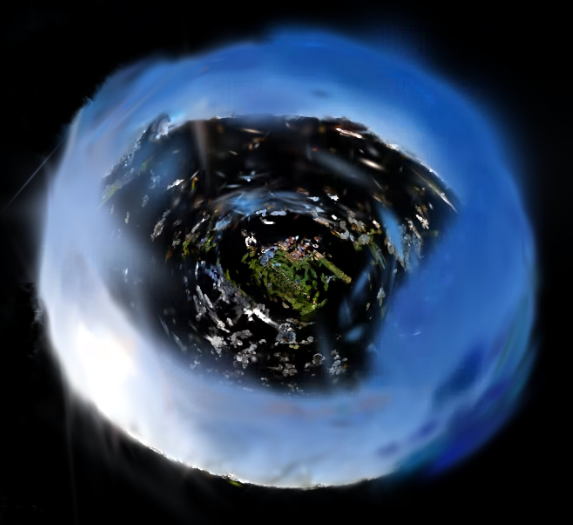}
            \caption{Our method (Two-Stage Two-Shells)}
            \label{fig:complete_gs_ours}
          \end{subfigure}
          \hspace{0.2cm}%
          \begin{subfigure}{0.31\linewidth}
            \includegraphics[width=\linewidth]{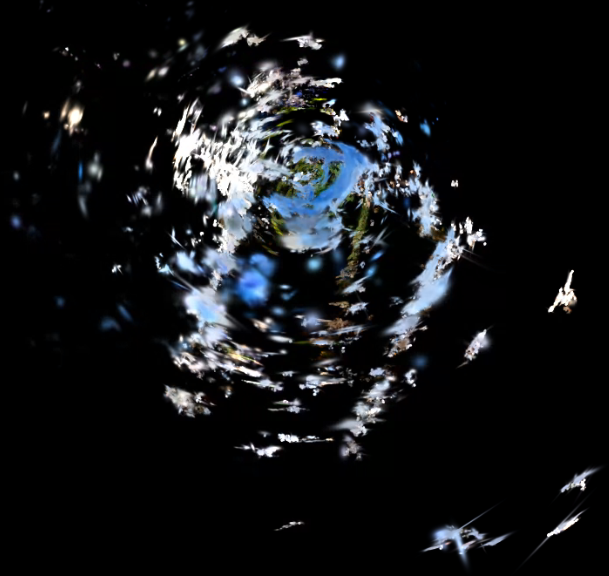}
             \caption{Original 3D Gaussian Splatting}      
            \label{fig:complete_gs_orig}
                     \end{subfigure}
          \hspace{0.2cm}%
          \begin{subfigure}{0.325\linewidth}
            \includegraphics[width=\linewidth]{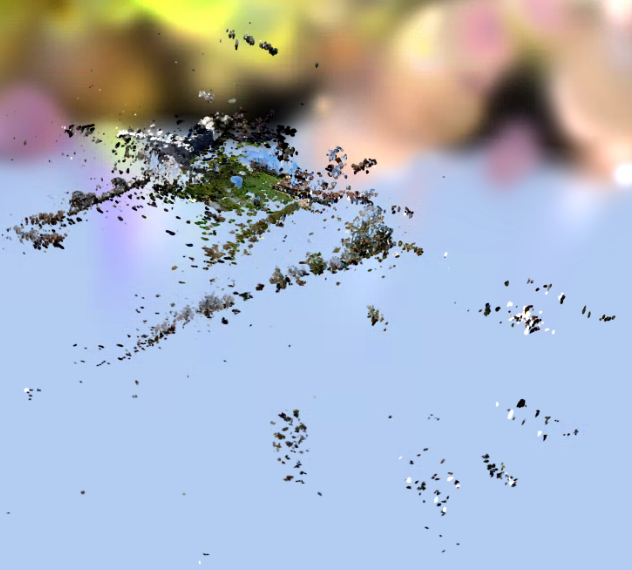}
             \caption{Hierarchical 3D Gaussian Splatting}  
            \label{fig:complete_gs_hie}
        \end{subfigure} 
          \caption{Comparison of distant views across methods. Our approach (a) preserves a spherical environment that is recognizable even from far away, while both the original (b) and hierarchical (c) 3D Gaussian Splatting tend to collapse into fragmented or distorted structures.}
          \label{fig:distance_three_methods}
        \end{figure*}

The reason why we have this clean result is well explained by looking from a distant view the distribution of the Gaussians generated by the different models, as done in \autoref{fig:distance_three_methods} for the \textit{Fields} dataset.
Our method results in a dense background representation and a limited number of small and well-localized Gaussians in the middle. The original GS code (b) creates a scattered set of Gaussians, which also determines the sky's appearance and distant objects placed at a short distance from the center.  
For the Hierarchical GS reconstruction (c), the external view mostly shows the low-resolution Gaussians in the background, which are rather coarse and may intersect the inner layers.

\subsection{Ablation Study}
\label{sec:abla}
We conducted an ablation study to evaluate the impact of the additional losses added in the first stage of the optimization pipeline. We also analyzed the effect of changing the random initialization of the points with radial depth within $R_0$ and $R_i$ with a different one based on the estimated depth.
We performed this study on the \textit{Person} dataset.

The results are summarized in \autoref{tab:ablation}.
The results show a substantial decrease in the quality of the novel view when we remove the new losses.
We expected this effect, as the goal of the method is to explicitly force the creation of a controlled background structure, avoiding the creation of floaters and artifacts.

When the planarity loss $L_{planarity}$ is removed, Gaussians tend to form radial spikes pointing toward the scene's center, producing a noisier reconstruction and a degradation in perceptual quality.

Turning off the shell loss $L_{shell}$ relaxes the geometrical constraint on the background Gaussians, resulting in a slightly increased SSIM compared to the planarity ablation. However, it introduces a highly blurry background and decreased visual fidelity.

Concerning the alternative method to initialize the outer shell's points, we observe that initializing these points following the distance maps, even if reasonable, results in slightly worse performance than using the default random choice.

\begin{table}[h]
\centering
\begin{tabular}{cll}
                                                                                &       &        \\
\multicolumn{1}{c|}{Person dataset}                                             &       &        \\ \hline
\multicolumn{1}{c|}{\multirow{3}{*}{Without Planarity Loss ($L_{planarity}$)}}                         & SSIM  & 0,508  \\
\multicolumn{1}{c|}{}                                                           & PSNR  & 14,291 \\
\multicolumn{1}{c|}{}                                                           & LPIPS & 0,497  \\ \hline
\multicolumn{1}{c|}{\multirow{3}{*}{Without Shell Loss ($L_{shell}$)}}                             & SSIM  & 0,531  \\
\multicolumn{1}{c|}{}                                                           & PSNR  & 14,783 \\
\multicolumn{1}{c|}{}                                                           & LPIPS & 0,504  \\ \hline

\multicolumn{1}{c|}{\multirow{3}{*}{All losses }}                             & SSIM  & 0,831  \\
\multicolumn{1}{c|}{}                                                           & PSNR  & 29,242 \\
\multicolumn{1}{c|}{}                                                           & LPIPS & 0,230  \\ \hline

\multicolumn{1}{c|}{\multirow{3}{*}{\parbox{0.50\linewidth}{All losses with distance-based initialization}}} & SSIM  & 0,826  \\
\multicolumn{1}{c|}{}                                                           & PSNR  & 29,001 \\
\multicolumn{1}{c|}{}                                                           & LPIPS & 0,237 
\end{tabular}
\caption{Ablation Study on the \textit{Person} dataset.}
\label{tab:ablation}
\end{table}

\subsection{Environment Maps' generation}
A practical, additional outcome of our approach, is the possibility of generating environment maps from the reconstructed background. 
The procedure is straightforward: once we complete the first optimization stage, we can use the Gaussian Splat rendering pipeline to render a cubic or spherical map on cameras placed inside the spherical shell. Depending on the distribution of the input cameras, the map will have empty areas, which can be easily filled with inpainting tools if required by the applications. 

\begin{figure}[h]
    \centering
    \includegraphics[width=1\linewidth]{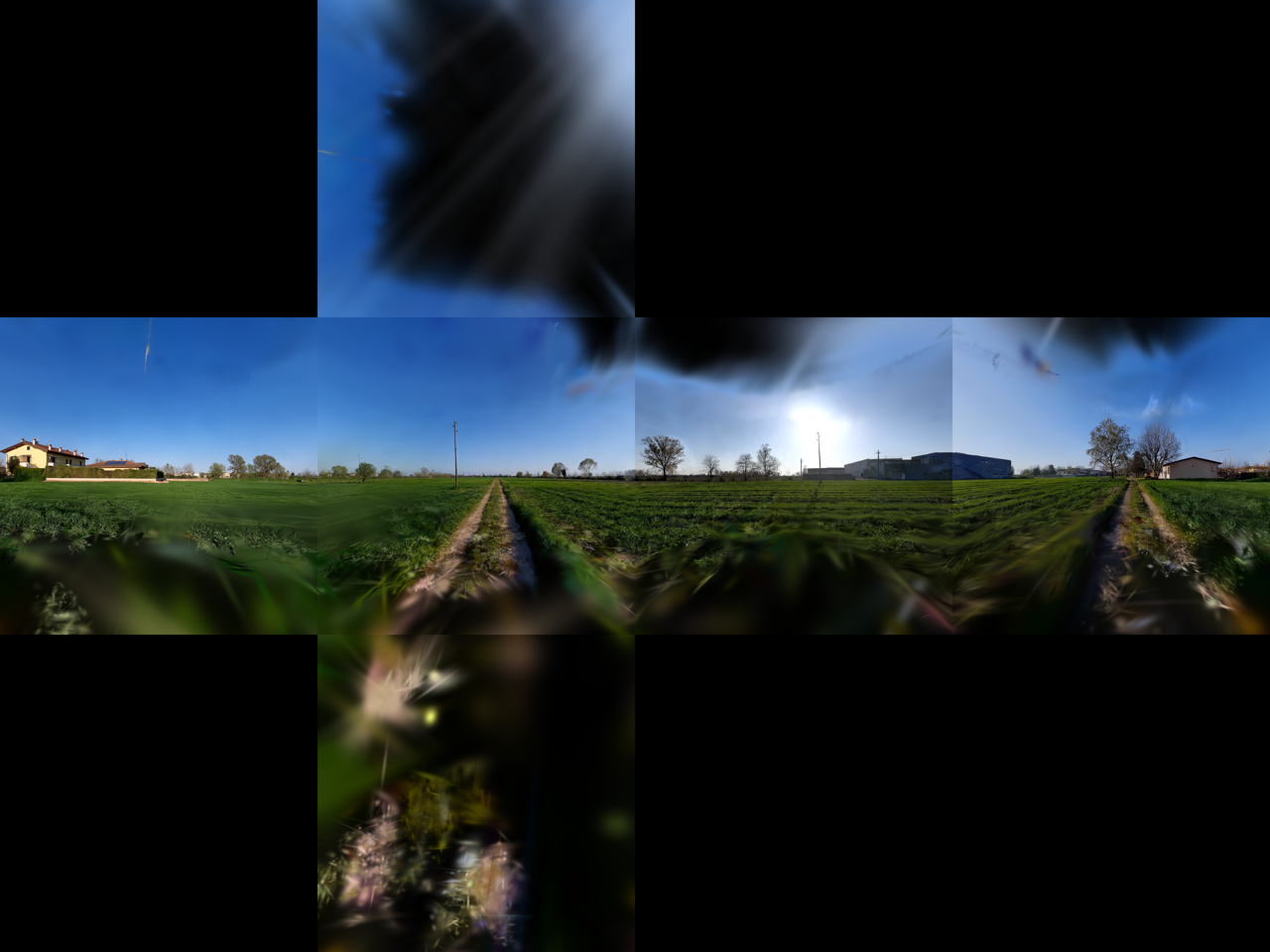}
    \caption{Cube map obtained from the first stage of the pipeline, representing the background of the scene in the \textit{Field} dataset.}
    \label{fig:envi}
\end{figure}

\autoref{fig:envi} shows an example of cube map obtained from the \textit{Fields} dataset.
Our tool allows the creation of environment maps, cutting all the parts closer than the wanted threshold, effectively merging the contributions of the available views. The examples generated with our benchmark data are obviously LDR. Still, our approach could be easily extended to the case of HDR Gaussian frameworks \cite{cai2024hdr}, and the LDR environment maps can also be expanded to HDR using recently proposed learning-based frameworks \cite{wang2023glowgan,bemana2025bracket}.

\section{Discussion}
We presented a two-stage/two-shells framework for reconstructing Gaussian environments, particularly suited for outdoor environments with distant objects and sky.
Our method leverages a segmentation of the image spaces based on the remapping of metric depth maps onto a global reference system centered in the "navigation" area, where we want to be able to create realistic, novel, synthetic views.
By optimizing a first set of Gaussians with properly designed losses to represent the background and the distant objects (outer shell), we can avoid the creation of floaters in the second step, optimizing the nearby part of the scene (inner shell), and obtain cleaner results, with considerable improvements in the ability to reconstruct a realistic sky and distant objects.
We have also shown that the optimized outer shell can be exploited to generate environment maps that may be useful to simplify the rendering of the scene, replacing the more complex Gaussian environment, or, when expanded to HDR or computed in HDR, for scene relighting.

While the results seem quite promising, the method comes with limitations. 
First, the interactive setting of the background distance, which can be a valuable feature for creators, makes the result user-dependent. We could introduce simple heuristics to set it automatically; however, we plan to investigate a strategy to design the automatic initialization to optimize the quality of the novel view synthesis results simultaneously.

While the initialization with the dense depth maps allows the creation of dense spherical backgrounds, these background still have holes due to missing input data. However, we can remove easily them with widely available image inpainting solutions if the application require it. We plan to investigate the option of introducing an inpainting strategy directly implemented in our Gaussian spherical backgrounds. 

An important outcome of our analysis is also related to the fact that standardized metrics used to evaluate novel view synthesis cannot fully capture the amount of artifacts and their visual impact. This fact is a significant research problem, as the value of the effectiveness of the light fields encoding methods is often evaluated based only on these metrics. 

In future work, we plan to design specific methods to directly assess the immersivity of this kind of light field reconstructions within test VR environments with user studies. 

Another issue for the evaluation is the lack of benchmarks with challenging outdoor scenes where test viewpoints are sufficiently far from the training ones. We plan to create new datasets, possibly with synthetic data, uniformly sampling test views within the area where the user can navigate to have a fair evaluation of the ability of the methods to create an immersive virtual environment.

One more consideration involves the computational costs. Our two-pass optimization naturally increases the total training time compared to single-pass baselines. This trade-off, however, does not impact runtime performance or the final Gaussian model in memory. 
The runtime rendering performance is unaffected, as the final representation is a standard set of 3D Gaussians, and its speed depends on the total primitive count. Similarly, the final Gaussian size in memory and total parameter count remain comparable to other methods, as our approach does not generate a significantly higher number of Gaussians (as shown in \autoref{tab:metrics}).
We trade a longer training process for an improvement in visual quality at no extra cost during real-time rendering.

Moreover, to address the longer training time, we will also look into unifying our pipeline into a single pass. This would optimize both layers at once, cutting the overall training cost without sacrificing the advantages of our two-shell approach.

\bibliographystyle{acm} 
\bibliography{main}       


\end{document}